\documentclass[aoas]{imsart}

\RequirePackage{amsthm,amsmath,amsfonts,amssymb}
\RequirePackage[authoryear]{natbib}
\RequirePackage[colorlinks,citecolor=blue,urlcolor=blue]{hyperref}
\RequirePackage{graphicx}
\RequirePackage{bm}
\RequirePackage{dsfont}

\usepackage{algorithm2e}
\DontPrintSemicolon

\startlocaldefs
\theoremstyle{plain}

\theoremstyle{remark}

\numberwithin{equation}{section}

\endlocaldefs

\begin{document}

\begin{frontmatter}
\title{A Flexible Bayesian Clustering of Dynamic Subpopulations in Neural Spiking Activity}
\runtitle{Bayesian Clustering of Neural Spiking Activity}

\begin{aug}
\author[A]{\fnms{Ganchao}~\snm{Wei}\ead[label=e1]{ganchao.wei@uconn.edu}},
\author[B]{\fnms{Ian H.}~\snm{Stevenson}\ead[label=e2]{ian.stevenson@uconn.edu}}
\and
\author[A]{\fnms{Xiaojing}~\snm{Wang}\ead[label=e3]{xiaojing.wang@uconn.edu}}
\address[A]{Department of Statistics,
	University of Connecticut \printead[presep={,\ }]{e1,e3}}

\address[B]{Department of Psychological Sciences,
	University of Connecticut\printead[presep={,\ }]{e2}}

\end{aug}

\begin{abstract}

With advances in neural recording techniques, neuroscientists are now able to record the spiking activity of many hundreds of neurons simultaneously, and new statistical methods are needed to understand the structure of this large-scale neural population activity. Although previous work has tried to summarize neural activity within and between known populations by extracting low-dimensional latent factors, in many cases what determines a unique population may be unclear. Neurons differ in their anatomical location, but also, in their cell types and response properties. To identify populations directly related to neural activity, we develop a clustering method based on a mixture of dynamic Poisson factor analyzers (mixDPFA) model, with the number of clusters and dimension of latent factors for each cluster treated as unknown parameters. To analyze the proposed mixDPFA model, we propose a Markov chain Monte Carlo (MCMC) algorithm to efficiently sample its posterior distribution. Validating our proposed MCMC algorithm through simulations, we find that it can accurately recover the unknown parameters and the true clustering in the model, and is insensitive to the initial cluster assignments. We then apply the proposed mixDPFA model to multi-region experimental recordings, where we find that the proposed method can identify novel, reliable clusters of neurons based on their activity, and may, thus, be a useful tool for neural data analysis.
\end{abstract}


\end{frontmatter}

\section{Introduction}
Identifying types of neurons is a longstanding challenge in neuroscience \citep{Nelson2006,Bota2007,Zeng2022}. Structural features based on anatomical location, cell morphology, genomics, developmental history, and synaptic connectivity have all been proposed, as well as Bayesian approaches for integrating these features \citep{Jonas2015}. Functional or electrophysiological features, based on neural activity, have also been widely used to distinguish between “types” of neurons \citep{Nowak2003}, especially in cases where the research focus is on understanding how the brain represents and processes information. Cell “type”, broadly defined, represents a form of much-needed organization – there are tens of millions of neurons in the mammalian brain (86 billion in humans). Framing analyses in terms of neuron “types” is often necessary for accurate description of the diversity of observations within an experiment and for generalization across experiments. Systematic, large-scale examination of intracellular features \citep{Gouwens2019} and responses to external stimuli \citep{Baden2016,DeVries2020} have produced rich new descriptions of functional cell types. However, there is often substantial heterogeneity within single cell types \citep{Tripathy2015,Cembrowski2019}, and except for the broadest categorizations (anatomical location and excitatory/inhibitory function) there is not a universally agreed upon taxonomy. Here we consider the problem of how to identify clusters of neurons based on simultaneous recordings of their spiking activity.

With modern techniques such as the high-density probes \citep{Jun2017,Steinmetz2021,Marshall2022}, we can have large-scale multi-electrode recordings from hundreds to thousands of neurons across different anatomical regions. Many experiments in systems neuroscience use these observations as their primary measurements \citep{Stevenson2011}, and several recent models have been developed to extract shared latent structures from simultaneous neural recordings, assuming that the activity of all of the recorded neurons can be described through common low-dimensional latent states. These approaches have proven useful in summarizing and interpreting high-dimensional population activity. Inferred low-dimensional latent states can provide insight into the representation of task variables \citep{Churchland2012,Mante2013,Cunningham2014,Saxena2019} and dynamics of the population itself \citep{Vyas2020}. Many existing approaches are extensions of two basic models: the linear dynamical system (LDS) model \citep{Macke2011} and a Gaussian process factor analysis (GPFA) model \citep{Yu2009}. The LDS model is built on the state-space model and assumes latent factors evolve with linear dynamics. On the other hand, GPFA models the latent factors by non-parametric Gaussian processes. However, by assuming a single unified “population” rather than discrete clusters of neurons, these models may miss important structure if it occurs in only a subset of neurons. Several variants of these models have been implemented to analyze multiple neural populations and their interactions \citep{SEMEDO2019249, NEURIPS2020_aa1f5f73}, but generally, the total number of clusters and the cluster membership is not evaluated systematically.

\begin{figure}[h!]
	\centering
	\includegraphics[width=1\textwidth]{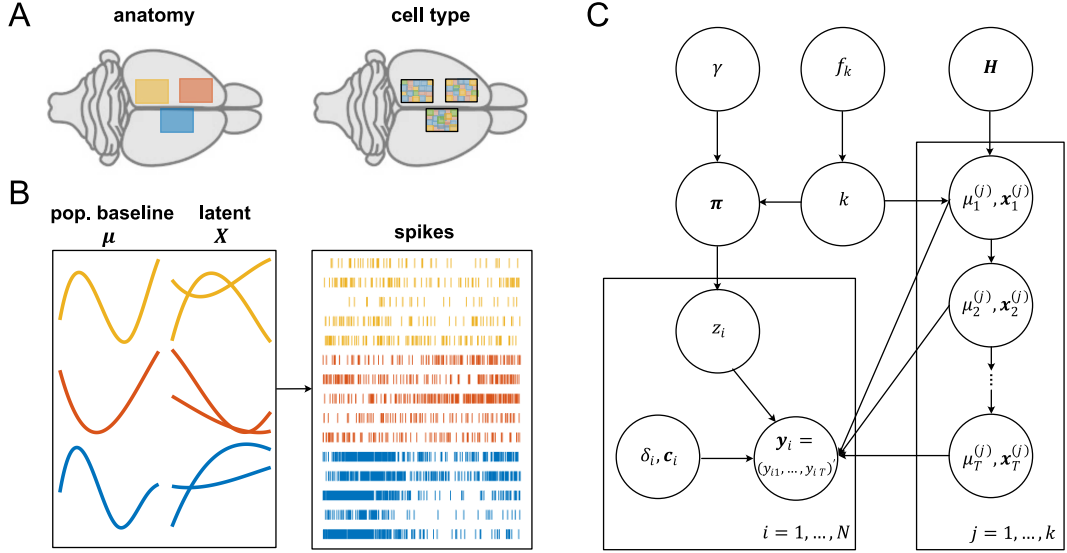}
	\caption{\textbf{Model overview. A.} The neural populations could be defined by anatomical regions (left) or by cell types (right). 
 \textbf{B.} The main goal is to cluster neurons according to their activity and extract functional grouping structure, based on spike train observations. The activity of each neuron is determined by a low dimensional latent state, specific to that neuron's cluster assignment (e.g. yellow, red, blue). \textbf{C.} Graphical representation of the key parts of the mixDPFA generative model. }
	\label{fig1}
\end{figure}

Neurons in different anatomical locations may interact with each other or receive common input from unobserved brain areas, sharing the same latent structure. On the other hand, neurons of different cell-types within the same brain area may be better described by distinct latent structures. From a functional point of view, neither the anatomical location nor cell type (Figure \ref{fig1}A) indicates which neurons should be grouped into the same populations. Previous researchers have used the idea of a “cell assembly” – a set of neurons with correlated activity \citep{Gerstein1989} – to conceptualize how a large population of neurons could consist of functionally relevant subpopulations or clusters. Cell assemblies have been hypothesized to arise as a result of learning and to provide distinct substrates for psychological processes \citep{Harris2005}. However, in many analyses, the cell assemblies are not treated as if they are distinct clusters with distinct psychological correlates \citep{Harris2003}. Identifying clusters of neurons and specific latent structure associated with each cluster may thus be a valuable tool for linking large-scale neural recordings to behavioral variables.

Motivated by the mixture of (Gaussian) factor analyzers (MFA, \citealp{Arminger1999,Ghahramani1996,Fokoue2003,Murphy2020}), which describes globally nonlinear data by combining a number of local factor analyzers, we group neurons here based on the latent factors (Figure \ref{fig1}B). By clustering neurons using a mixture of latent factor models, we generate a novel description of neural populations – with distinct clusters and distinct state variables for each cluster. By comparing the cluster assignments to other descriptions of cell types (e.g., anatomy, genomics, progenitors, electrophysiological features, synaptic connectivity), we may be able to better understand when and how these “types” act together in vivo, and how distinct neuronal types function (or do not function) as integrated cell assemblies. The estimated latent factors can then reflect “separable” patterns of covariation. Further, by comparing the trajectories of latent factors to stimulus or task variables, we may be able to better understand to what extent different task features are represented by different subpopulations.

A similar approach to the problem of clustering neurons by latent structures was previously developed using a mixture of Poisson linear dynamical systems (PLDS) model (mixPLDS, \citealt{NIPS2014_e8dfff46}). The mixPLDS model infers the subpopulations and latent factors using deterministic variational inference \citep{MAL-001,Jordan1999,pmlr-v28-emtiyazkhan13} and the model parameters are estimated by an Expectation Maximization algorithm \citep{Dempster1977}. Unlike MFA, the mixPLDS can capture temporal dependencies of neural activity as well as interactions between clusters over time. However, there are several limitations for mixPLDS: 1) it requires we predetermine the number of clusters and the dimension of latent vectors for each cluster,  and 2) the clustering results are sensitive to the initial cluster assignment.

Here we cluster the neurons by a mixture of dynamic Poisson factor analyzers (mixDPFA). The mixDPFA model takes the advantages of both Poisson factor analysis (FA) and PLDS as well as includes both a population baseline and baselines for individual neurons. 
Both the number of clusters and latent factor dimensions are treated as unknown parameters in our proposed mixDPFA model, and we sample the posterior distributions using a Markov Chain Monte Carlo (MCMC) algorithm. To sample the latent factors, we develop an efficient Metropolis-Hasting algorithm using the P\'olya-Gamma data augmentation technique \citep{doi:10.1080/01621459.2013.829001}. Further, to sample the number of latent factors for each cluster, we employ a birth-and-death MCMC (BDMCMC) algorithm \citep{10.1214/aos/1016120364}. Moreover, to facilitate the sampling of the cluster indices, we use the partition-based algorithm developed in mixture of finite mixtures (MFM) model by \cite{Miller2018}. To improve mixing and accelerate our computation speed of our proposed mixDPFA model, we evaluate the approximated marginalized likelihood when sampling cluster indices and the latent dimension in each cluster, by using the Poisson-Gamma conjugacy. After validating the proposed model with simulated data, we apply it to analyze multi-region experimental recordings from behaving mice. Overall, the proposed method provides a way to efficiently cluster neurons into populations based on their activity.

The rest of the paper is structured as follows. In Section \ref{method} we introduce our mixDPFA model. We then propose an efficient MCMC algorithm to sample posterior distributions of mixDPFA parameters in Section \ref{infMethod}. We present an analysis of synthetic datasets in Section \ref{sim}, while in Section \ref{application} we apply the proposed clustering method to analyze experimental data (large-scale spike recordings from the Visual Coding Neuropixels Dataset from the Allen Institute for Brain Science, \citet{Siegle2021}). Finally, in Section \ref{discussion}, we conclude with some final remarks and highlight some potential extensions of our current model for future research.

\section{Mixture of Dynamic Poisson Factor Analyzers Model}
\label{method}
Here we propose a mixture of dynamic Poisson factor analyzers (mixDPFA) to cluster neurons based on multi-population latent structures. After introducing the model, we will provide the prior specification and derive the joint posterior distribution for unknown parameters in the model. 

\subsection{Introduction of the Proposed mixDPFA}

 To make the explanation of the proposed model easier, first,  we just describe the single population dynamic Poisson factor analyzer (DPFA) model with a given cluster, and then we introduce how to specify the priors on the number of clusters and clarify how we use the mixture of finite mixtures (MFM) for the DPFA to cluster neural activity.

\subsubsection{Dynamic Poisson Factor Analyzer}
\label{DPFA}
Denote the observed spike count of neuron $i \in \{ 1,\ldots,N\}$ at
time bin $t \in \{ 1,\ldots,T\}$ as
$y_{it}$ (a non-negative integer) and the cluster indicator of neuron $i$ as $z_{i}$. Motivated by the nature of neural activity and the former PLDS model \citep{Macke2011}, we propose a dynamic Poisson factor analysis model by adding individual baselines $\delta_i$. The proposed model is a combination of PLDS and Poisson factor analysis model, which includes both population baseline and individual baselines. Assume neuron $i$ belongs to the $j$-th cluster (i.e., $z_i=j$), and its spiking activity is independently Poisson distributed, conditional on the
low-dimensional latent state $\bm{x}_{t}^{(j)} \in \mathbb{R}^{p_j}$, where $p_j$ indicates the dimension of latent state $\bm{x}_{t}^{(j)}$ on the $j$th cluster and population baseline $\mu_t^{(j)}$, then the observation equation for our model is as follows:
\begin{eqnarray}
	\label{genModel}
	\begin{split}
		y_{it} &\sim Poi(\lambda_{it}), \\
		\log\lambda_{it} &= \delta_i + \mu_t^{(j)} + \bm{c}'_i\bm{x}^{(j)}_t.
	\end{split}
\end{eqnarray}
Notice in the observation equation \eqref{genModel}, the neuron-specific baseline $\delta_i$ is a constant across time for the $i$th neuron and unrelated to the cluster assignment. Further, we assume the population baseline $\mu_t^{(j)}$ and the latent state $\bm{x}^{(j)}_t$ in the observation equation \eqref{genModel} evolve linearly over time with Gaussian noises as shown in the system equation below, 
\begin{eqnarray}
	\label{stateEquation}
	\begin{split}
		\mu_{t+1}^{(j)}=g^{(j)}+h^{(j)}\mu_t^{(j)} +\epsilon_t^{(j)},\\ \bm{x}_{t+1}^{(j)}=\bm{b}^{(j)}+\bm{A}^{(j)}\bm{x}_t^{(j)}+\bm \eta_t^{(j)},
	\end{split}
\end{eqnarray}
where $\epsilon_t^{(j)}\sim \mathcal{N}(0,(\sigma^{2})^{(j)})$ and $\bm \eta_t^{(j)} \sim \mathcal{N}_{p_j} (\bm 0, \bm{Q}^{(j)})$ with $\bm{Q}^{(j)}$ as an unknown covariance matrix in the $j$th cluster. The full observation model allows us to account for differences between the average firing rates of different clusters, since $\mu_{t}^{(j)}$ acts as a "gain" on all neurons in the cluster, and differences in the dynamics or firing rate evolution of different clusters (via $g^{(j)}$, $h^{(j)}$, $\bm{b}^{(j)}$, and $\bm{A}^{(j)}$).

If we denote $\bm{y}_{i} =  (y_{i1},\ldots,y_{iT})'$, $\bm{\lambda}_{i} =  (\lambda_{i1},\ldots,\lambda_{iT})'$, $\bm{\mu}^{(j)} = (\mu^{(j)}_1,\ldots,\mu^{(j)}_T)'$, $\bm{X}^{(j)} = (\bm{x}^{(j)}_1,\ldots,\bm{x}^{(j)}_T)'$ and $\bm 1_T$ to be a $T\times 1$ column vector with each element being 1,  the proposed observation equation can be rewritten in a matrix notation as below, 
\begin{align}
	\begin{split}
		\bm{y}_i &\sim Poi(\bm{\lambda}_i),\\
		\log\bm{\lambda}_i &= \delta_i\bm{1}_T + \bm{\mu}^{(j)}+ \bm{X}^{(j)}\bm{c}_i.\label{eq_matrix}
	\end{split}
\end{align}
Generally, a factor model is consistent only when $T/N \rightarrow 0$ \citep{Johnstone2009}, but this is often not the case for most neural spike data. However, when we assume linear dynamics on $\bm{\mu}^{(j)}$ and $\bm{X}^{(j)}$, it resolves the consistency issue. Further, as known in a FA model, when $p_j>1$, the model is only identifiable up to orthogonal rotation on $\bm{X}^{(j)}$, with $\bm{c}_i\sim \mathcal{N}_{p_j}(\bm{0},\bm{I}_{p_j})$ with $\bm I_{p_j}$ indicating an identity matrix with $p_j$ dimension. By including an individual baseline $\delta_i\bm{1}_T$ in  \eqref{eq_matrix}, it also makes the proposed model invariant to translation of $\bm{\mu}^{(j)}$ and $\bm{X}^{(j)}$. To make the model identifiable and encourage clustering based on the trajectories of latent factors, we assume $\bm{A}^{(j)}$ and $\bm{Q}^{(j)}$ are diagonal following the suggestions from \citet{Pena2004} and \citet{Lopes2008}, i.e., $\bm{A}^{(j)} = diag(a_1^{(j)},\ldots, a_{p_j}^{(j)})$ and $\bm{Q}^{(j)} = diag(q_1^{(j)},\ldots, q_{p_j}^{(j)})$, and moreover, we assume $\sum_{t=1}^{T}\mu_t^{(j)} = 0$ and $\sum_{t=1}^{T}\bm{x}^{(j)}_t = \bm{0}$.

Given the parameters of the $j$-th cluster $\bm{\theta}^{(j)} = \{p_j, \bm{\mu}^{(j)}, \bm{X}^{(j)}, h^{(j)}, g^{(j)}, (\sigma^2)^{(j)},\bm{A}^{(j)},\allowbreak \bm{b}^{(j)},\allowbreak \bm{Q}^{(j)}\}$, then the spike counts of neuron $i$ are generated by the DPFA model as $[\bm{y}_i \mid z_i=j]\sim DPFA(\delta_i, \bm{c}_i, \bm{\theta}^{(j)})$. To facilitate the Bayesian computation for this complex model, we need to impose priors $\bm{H}$ on $\bm{\theta}^{(j)}$, which are introduced in Section \ref{infer}.

\subsubsection{Clustering by Mixture of Finite Mixtures Model}
\label{MFM}
When the population labels $z_{i}$ are unknown, we cluster the neurons by a mixture of DPFA (mixDPFA). We assume the number of neural populations $k$ is finite but unknown and thus we need to put some priors on it. To make the Bayesian computation more efficient, we utilize the idea from the mixture of finite mixtures (MFM, \cite{Miller2018}) model, by assigning the priors for the clusters
\begin{equation}
	\label{MFMM}
	\begin{aligned}
		k &\sim f_k, &&f_k \: \text{is a p.m.f. on}
		\{1,2,\ldots\},\\
		\bm{\pi}=(\pi_1,\ldots,\pi_k) &\sim Dir_k(\gamma, \ldots,\gamma) &&\text{given}\: k, \\
		z_1,\ldots,z_N&\stackrel{i.i.d.}{\sim}\bm{\pi} &&\text{given}\: \bm{\pi}, \\
		\bm{\theta}^{(1)},\ldots,\bm{\theta}^{(k)}&\stackrel{i.i.d.}{\sim}\bm{H} &&\text{given}\: k,\\
		\bm{y}_i = (y_{i1},\ldots,y_{iT})' &\sim {\rm DPFA}(\delta_i,\bm{c}_i,\bm{\theta}^{(z_i)}) &&\text{given} \ \delta_i, \bm c_i, \bm \theta^{(z_i)}, z_i, \forall i=1,\ldots,N,
	\end{aligned}
\end{equation}
where ${\rm p.m.f}$ denotes the probability mass function. 
By using the MFM, we can integrate the field knowledge about the number of neural populations into our analysis.

Another way to handle the unknown number of clusters is to use the Dirichlet process model (DPM). However, MFM may be more appropriate than DPM conceptually in our case, since the number of neural "populations" is unknown but finite. Additionally, MFM produces more concentrated, evenly dispersed clusters (see \cite{Miller2018} for detailed discussion). The key parts of the generative model is summarized in a graphical form shown in Figure \ref{fig1}C.

\subsection{Prior Specifications}
\label{prior}
In the observation equation \eqref{genModel}, we assume 1) the dimension of the latent factor $p_j$ follows a truncated Poisson prior (set maximum of $p_j$ be 20, as usually 2-4 latent factors are enough for both data description and model interpretation) with hyperparameter $\alpha = 2$, i.e., $P(p_j)\propto \alpha^{p_j}/ p_j !$, for $p_j = 1,..., 20$ \citep{10.1214/aos/1016120364,Fokoue2003}; 2) the initial population baseline and latent factor at $t=1$ follow $\mu_1^{(j)} \sim \mathcal{N}(0, 1)$ and $\bm{x}_1^{(j)} \sim \mathcal{N}_{p_j}(\bm{0}, \bm{I}_{p_j})$; and 3) the neuron-specific baseline $\delta_i \sim \mathcal{N}(0,1)$ and factor loading $\bm{c}_i \sim \mathcal{N}_{p_j}(\bm{0}, \bm{I}_{p_j})$. For  parameters in the system equation \eqref{stateEquation}, since we assume $\bm{A}^{(j)}$ and $\bm{Q}^{(j)}$ are diagonal, the priors and updates for $a_m^{(j)}$, $b_m^{(j)}$ and $q_m^{(j)}$ are similar to those for $h^{(j)}$, $g^{(j)}$ and $(\sigma^{2})^{(j)}$ for each of $m=1,\ldots,p_j$. These priors are set as $(\sigma^{2})^{(j)} \sim IG\left(\nu_0/2, \nu_0\sigma_0^2/2\right)$ (with the same type of priors on $q_m^{(j)}$), where $IG$ denotes the inverse-gamma distribution and $[( g^{(j)}, h^{(j)})'\mid (\sigma^{2})^{(j)}] \sim \mathcal{N}(\bm{\tau}_0, (\sigma^{2})^{(j)}\bm{I}_2)$ (with the same type of priors on $[( b_m^{(j)}, a_m^{(j)})'\mid (q_m)^{(j)}]$), with $\nu_0 = 1$, $\sigma^2_0 = 0.01$ and $\bm{\tau}_0 = (0,1)'$. In Equation \eqref{MFMM}, to encourage the number of clusters to be small, we assume $k\sim Geometric(\nu)$, with its density defined as $f_k(k|\nu) = (1-\nu)^{k-1}\nu$ for $k = 1,2,\ldots$,  and let $\gamma = 1$ for $\bm{\pi} \sim Dir_k(\gamma,\ldots,\gamma)$. The choice of $\nu$ depends on applications.

\subsection{Likelihood and Posterior Distribution}
\label{infer}

According to the observation equation (\ref{genModel}), given the cluster index $z_i = j$ for neuron $i$, the likelihood function for $(\delta_i, \bm{c}_i, \bm{\theta}^{(j)})$ is:
\begin{equation}
	\label{likelihood}
	P(\bm{y}_i|\delta_i, \bm{c}_i,\bm{\theta}^{(j)}) = \prod_{t=1}^{T}e^{-\lambda_{it}}\lambda_{it}^{y_{it}}/y_{it}!. 
\end{equation}
where, $\lambda_{it} = \exp(\delta_i + \mu_t^{(j)} + \bm{c}'_i\bm{x}_t^{(j)})$. 
On the other hand, by the system equation \eqref{stateEquation} and the specified prior for $\mu_1^{(j)}$ and $\bm{x}_1^{(j)}$, 
\begin{align*}
P(\widetilde{\bm{X}}^{(j)}\mid \bm{\xi}^{(j)}) &\propto \exp{\left(-\frac{1}{2}\| \widetilde{\bm{x}}_1^{(j)}\|_2^2\right)}\prod_{t=2}^{T}\exp{\left(-\frac{1}{2}\bm{s'}^{(j)}_t| \widetilde{\bm{Q}}^{(j)}|^{-1}\bm{s}_t^{(j)}\right)},
\end{align*}
where $\bm{s}_t^{(j)} = \widetilde{\bm{x}}_t^{(j)} - \widetilde{\bm{A}}^{(j)}\widetilde{\bm{x}}_{t-1}^{(j)} - \widetilde{\bm{b}}^{(j)}$, $\bm{\xi}^{(j)} = \{\widetilde{\bm{b}}^{(j)},\widetilde{\bm{A}}^{(j)},\widetilde{\bm{Q}}^{(j)}\}$, $\widetilde{\bm{X}}^{(j)} = \left(\bm{\mu}^{(j)}, \bm{X}^{(j)}\right)$, $\widetilde{\bm{x}}_t^{(j)} = \left(\mu_t^{(j)}, \bm{x}'^{(j)}_t\right)'$, $\widetilde{\bm{A}}^{(j)} = diag(h^{(j)},\allowbreak \bm{A}^{(j)})$, $\widetilde{\bm{b}}^{(j)} = \left(g^{(j)}, \bm{b}'^{(j)}_t\right)'$ and $\widetilde{\bm{Q}}^{(j)} = diag((\sigma^{2})^{(j)}, \bm{Q}^{(j)})$.

Therefore, the joint posterior distribution of parameters in the mixDPFA can be represented by a countably infinite summation (following from \citet{10.1214/21-BA1294}) as follows:
\begin{equation}
\label{FullPost}
\begin{aligned}
    P(\{\delta_i\}_{i=1}^N,\{\bm{c}_i\}_{i=1}^N,\{\bm{\theta}^{(j)}\}\mid \{\bm{y}_i\}_{i=1}^N) &\propto \sum_{k=1}^{\infty} f_k(k)\left[\prod_{i=1}^{N}P(\bm{y}_i\mid k)P(\delta_i)P(\bm{c}_i)\right]\\
    &\times \left[\prod_{j=1}^{k}P(p_j)P(\widetilde{\bm{X}}^{(j)}\mid \bm{\xi}^{(j)})P(\bm{\xi}^{(j)})\right],
\end{aligned}
\end{equation}
where we somewhat abuse the notation by defining $P(\bm{y}_i\mid k) = \sum_{j=1}^{k} P(z_i=j) P(\bm{y}_i\mid \delta_i, \bm{c}_i, \bm{\theta}^{(z_i)}, z_i=j)=\sum_{j=1}^{k} \pi_k P(\bm{y}_i\mid \delta_i, \bm{c}_i, \bm{\theta}^{(j)})$, noticing that $P(\bm \pi)\propto 1$ and $P(\bm{\xi}^{(j)})$ denoting the priors we have assigned on $g^{(j)}, h^{(j)}, \bm b^{(j)}, \bm A^{(j)}, (\sigma^2)^{(j)}$ and ${\bm Q}^{(j)}$. 

\section{Statistical Inference for the mixDPFA Model}
\label{infMethod}
The joint posterior distribution shown in Equation \eqref{FullPost} is not in closed form, thus we have to resort to an MCMC algorithm to sample it (see more details in the Appendix \ref{MCMC}). We will briefly introduce our key steps in this section. In each iteration, there are three key steps: 1) sample the model parameters assuming the cluster indices and latent dimension for each cluster are given, 2) find the dimension of latent states for each cluster and 3) draw the cluster indices given the model parameters. Step 1) can be further decomposed into three blocks, i.e., sampling of a) population baseline $\bm{\mu}^{(j)}$ and latent factors $\bm{X}^{(j)}$ for each cluster, b) cluster-independent $\delta_i$ and $\bm{c}_i$, and c) parameters for linear dynamics in each cluster $\{h^{(j)}, g^{(j)}, (\sigma^2)^{(j)}, \bm{A}^{(j)}, \bm{b}^{(j)}, \bm{Q}^{(j)}\}$.

\subsection{Key Sampling Steps in the mixDPFA Model}
When sampling the model parameters $\bm \theta^{(z_i)}$ given $z_i=j$ and latent dimension for each cluster, the full conditional distribution of the latent state $\bm{X}^{(j)}$ and population baseline $\bm{\mu}^{(j)}$ is equivalent to the posterior distribution of the dynamic Poisson generalized linear model, which has no closed form (see Equation \eqref{eqn_cd_xu} in Appendix \ref{xmu}). Thus, we sample the posteriors by a P\'olya-Gamma (PG) data augmentation approach \citep{Windle2013,doi:10.1080/01621459.2013.829001} with an additional Metropolis-Hastings (MH) step \citep{doi:10.1063/1.1699114, 10.1093/biomet/57.1.97}. Although the Poisson observations don't follow the PG augmentation scheme directly, we can approximate the Poisson distribution by a negative binomial (NB) distribution. After approximating Poisson with NB, we augment the model by introducing additional PG variables, and then we sample the vector $\bm{X}^{(j)}$ and $\bm{\mu}^{(j)}$ as a whole by using the forward-filtering-backward-sampling (FFBS) algorithm. To ensure the samples are exactly from the full conditional distributions (c.f., Equation \eqref{eqn_cd_xu} in Appendix \ref{xmu}), we use samples from the FFBS algorithm as a proposal and employ a MH step to reject or accept this proposal (more details can be found in Appendix \ref{xmu} and Algorithm \ref{algo}). The dispersion parameter ($r_{it}$ in Algorithm \ref{algo}) for the NB approximation can be used as a tuning parameter to make MH achieve desirable acceptance rate (in the experiments here, we aim for an acceptance rate of 0.4-0.5).

Once we update the latent state $\bm{X}^{(j)}$ and population baseline $\bm{\mu}^{(j)}$, we then update the cluster-independent parameters $\delta_i$ and $\bm{c}_i$ and the parameters governing the linear dynamics for each cluster, i.e., $h^{(j)}, g^{(j)}, (\sigma^2)^{(j)}, \bm{A}^{(j)}, \bm{b}^{(j)}$ and $\bm{Q}^{(j)}$. The sampling of $\delta_i$ and $\bm{c}_i$ (Appendix \ref{deltac}) from the full conditional distribution is equivalent to sampling the parameters of a Poisson regression, and we use Hamiltonian Monte-Carlo (HMC, \citet{Duane1987,Neal1994}) to update them. The parameters governing linear dynamics (Appendix \ref{dynamics}) are updated using a Gibbs sampler because of the conjugacy between priors and full conditional distributions.

Next, to sample the number of latent factors $p_j$ in each cluster, we use birth-and-death MCMC (BDMCMC) (c.f., \cite{10.1214/aos/1016120364} and \cite{Fokoue2003}), which requires very little mathematical sophistication and is easy for interpretation. In the previous research, there are ways to put a multiplicative Gamma process prior \citep{Bhattacharya2011}, multiplicative exponential process prior \citep{pmlr-v51-wang16e}, Beta process prior \citep{10.1145/1553374.1553474,5559508} or Indian Buffet process prior\citep{10.1007/978-3-540-74494-8_48,10.1214/10-AOAS435,doi:10.1080/01621459.2015.1100620} on the loading matrix in the Gaussian factor analysis model. Although these methods may be better than BDMCMC in some cases, it is not easy to implement them in our case. Since putting prior on $\bm{c}_i$ makes evaluation of marginal likelihood difficult, while putting prior on $\bm{X}^{(j)}$ will break the assumed linear dynamics. But motivated by these methods, it may be possible to put prior on linear dynamics ($\bm{A}^{(j)}$, $\bm{b}^{(j)}$ and $\bm{Q}^{(j)}$) to encourage shrinkage in $\bm{X}^{(j)}$. To efficiently simulate a birth-death Markov point process to estimate $p_j$ for the high dimensional (i.e., large $T$) mixDPFA model, we evaluate the marginalized likelihood by integrating out the neuron-specific $\bm{c}_i$ in Equation \eqref{likelihood}, that is, 
\begin{equation}
	M_{\bm{\theta}^{(j)}}(\bm{y}_i) = P(\bm{y}_i|\bm{\theta}^{(j)}, \delta_i) = \int P(\bm{y}_i|\bm{\theta}^{(j)},\delta_i,\bm{c}_i)P(\bm{c}_i)\,d\bm{c}_i,\label{eq_marginal}
\end{equation}
which is the marginalized likelihood of neuron $i$ in cluster $j$.  This marginalized likelihood has no closed form, and would be computationally intensive to approximate when iterating over all clusters and latent dimension. Since our primary goal in this step is to update the latent dimension $p_j$ not necessarily to calculate the exact marginalized likelihood, here we approximate the marginalized likelihood (Equation \eqref{eq_marginal}) by utilizing a Poisson-Gamma conjugacy. This approach has been previously utilized to approximate posteriors \citep{El-Sayyad1973} and predictive distributions \citep{Chan2009}. In our situation, since $\bm{c}_i\sim \mathcal{N}_{p_i}(\bm{0},\bm{I}_{p_i})$, we have $\lambda_{it} = \exp(\delta_i + \mu_t^{(j)} + \bm{c}'_i\bm{x}^{(j)}_t)\sim lognormal(\delta_i + \mu_t^{(j)}, \bm{x}'^{(j)}_t\bm{x}^{(j)}_t)$, and then we can approximate this lognormal distribution by a gamma distribution, i.e., assume $\lambda_{it}$ follows $Gamma(a_{it}, b_{it})$ with $a_{it} = (\bm{x}'^{(j)}_t\bm{x}^{(j)}_t)^{-1}$ as a shape parameter and $b_{it} = \bm{x}'^{(j)}_t\bm{x}^{(j)}_t\cdot e^{\delta_i + \mu_t^{(j)}}$ as a scale parameter. Thus, by the conjugate property with Poisson and Gamma random variables, we have
\begin{equation}
	\label{approx_margLike}
	P(y_{it}\mid \bm{\theta}^{(j)}, \delta_i) = \int P(y_{it}\mid \lambda_{it})P(\lambda_{it})\,d\lambda_{it}\approx NB(y_{it}\mid \nu_{it}, p_{it}),
\end{equation}
where $NB(y_{it}\mid \cdot, \cdot)$ denotes that $y_{it}$ follows a negative binomial distribution with  $\nu_{it} = a_{it}$ and $p_{it} = 1/(1 + b_{it})$ as its parameters. Since the observations are conditionally independent, that is $P(\bm{y}_i\mid \bm{\theta}^{(j)}, \delta_i)=\prod_{t=1}^{T}P(y_{it}\mid \bm{\theta}^{(j)}, \delta_i)$, we then have a closed-form approximation for Equation \eqref{eq_marginal}. This provides a computationally inexpensive approach to sample latent dimension $p_j$ for each cluster. The detailed sampling steps can be found in Appendix \ref{p} and Algorithm \ref{bdmcmc}.

After updating $\delta_i$, $\bm{c}_i$ and $\bm{\theta}^{(j)}$, the cluster indices  $z_i$'s are then sampled using an approach that is analogous to the partition-based algorithm in Dirichlet process mixtures (DPM, \citet{Neal2000}), with "split-merge" Metropolis-Hasting steps \citep{10.1214/07-BA219,doi:10.1198/1061860043001}
as in \citep{Miller2018}. The samples for estimated number of neural population $k$ is obtained by calculating the number of unique $\{z_i\}_{i=1}^N$. Although the exact posterior distribution of $k$ can be calculated post-hoc as in \citet{Miller2018}, we don't need to explicitly infer it in our algorithm since our MCMC sampling scheme is based on the joint posterior by marginalizing out $k$. 
The empirical partitions of neurons is easily obtained from the MCMC results and it is practically more interesting, thus we do not infer the exact posterior of $k$ here. 
When sampling the clustering indices $z_i$'s in the high dimensional time series data with large $T$, we will also just evaluate the approximate marginalized likelihood of Equation \eqref{eq_marginal} since we need to evaluate Equation \eqref{eq_marginal} many times by iterating over all potential clusters for each neuron. The same computationally efficient approximation (Equation \eqref{approx_margLike}) is used to facilitate this sampling of cluster indices $z_i$'s.  Such computational efficiency is important in applications to real data, especially when the numbers of neurons and potential clusters are large. Detailed sampling strategies for $z_i$'s can be founded in Appendix \ref{zi}. 

MATLAB code for the mixDPFA model is available at \url{https://github.com/weigcdsb/MFM_DPFA_clean}, and additional details for MCMC sampling can be found in Appendix \ref{MCMC}..

\subsection{Contributions on the Analysis of mixDPFA}

The MCMC approach developed here for the mixDPFA model includes several improvements on previous latent variable models of neural spiking activity. First, unlike the algorithm for the mixture of PLDS (mixPLDS,  \citet{NIPS2014_e8dfff46}), we do not need to pre-specify the number of clusters and latent dimensions; instead, all these quantities can now be inferred from the data using our proposed mixDPFA model. Second, the sampling of $\bm{X}^{(j)}$ and $\bm{\mu}^{(j)}$ in the previous MFA or PLDS is not trivial. Previous work in \citet{pmlr-v54-linderman17a} and \citet{NIPS2016_708f3cf8} approximated the Poisson distributed observations by the NB distribution with large dispersion parameters, and used Gibbs sampling with PG augmentation and FFBS. However, this approximation may introduce bias in posterior inference and the convergence of the Gibbs sampler may suffer from the high auto-correlations of latent variables. By adding one more MH step, we ensure the samples are from the exact posterior and also achieve better mixing of latent variables. In the MH step, the dispersion parameter $r_{it}$ for NB distribution in Algorithm \ref{algo} becomes a tuning parameter, which can balance the acceptance rate and autocorrelation. Finally, when doing the clustering, we need to evaluate the likelihood for neurons under each cluster. If we simply use the data-augmentation or imputation-posterior algorithm as in the MCMC proposed for Gaussian MFA \citep{Fokoue2003}, i.e., sampling $\bm{c}_i$ directly and evaluating the likelihood (Equation \eqref{likelihood}), the chain has poor mixing and stops after a few iterations because of the high dimensionality. The heavy dependency on the starting point for the previous algorithm of mixPLDS \citep{NIPS2014_e8dfff46} may suggest a similar problem. Here we provide a computationally efficient solution by evaluating a closed-form approximation of the marginalized likelihood instead. The same approximation technique allows the dimension of the latent factors to be sampled efficiently.

\section{Simulations}
\label{sim}
To validate and illustrate the proposed clustering method, we generate simulated neural data directly from the observation equation \eqref{genModel}. To generate $\bm{\mu}^{(j)}$ and $\bm{X}^{(j)}$, we first randomly pick up 10 to 35 points evenly spacing across recording length $T$, and the corresponding value for each point is generated from $\mathcal{N}(0,0.5^2)$. Then, the whole trajectories of $\bm{\mu}^{(j)}$ and $\bm{X}^{(j)}$ are generated by spline interpolation passing through these points. We simulate 10 clusters with 5 neurons in each (therefore there are $N=50$ neurons in total), with recording length $T=1000$ and $p_j=2$ dimensional latent factors for each cluster. In each cluster, $\delta_i\sim \mathcal{N}(0, 0.5^2)$, $\bm{c}_i\sim \mathcal{N}_2(\bm{0},\bm{I}_2)$.  The spiking activity of these simulated neurons are generated 50 times with different random seeds, using the same set of parameters.

\begin{figure}[h!]
	\centering
	\includegraphics[width=1\textwidth]{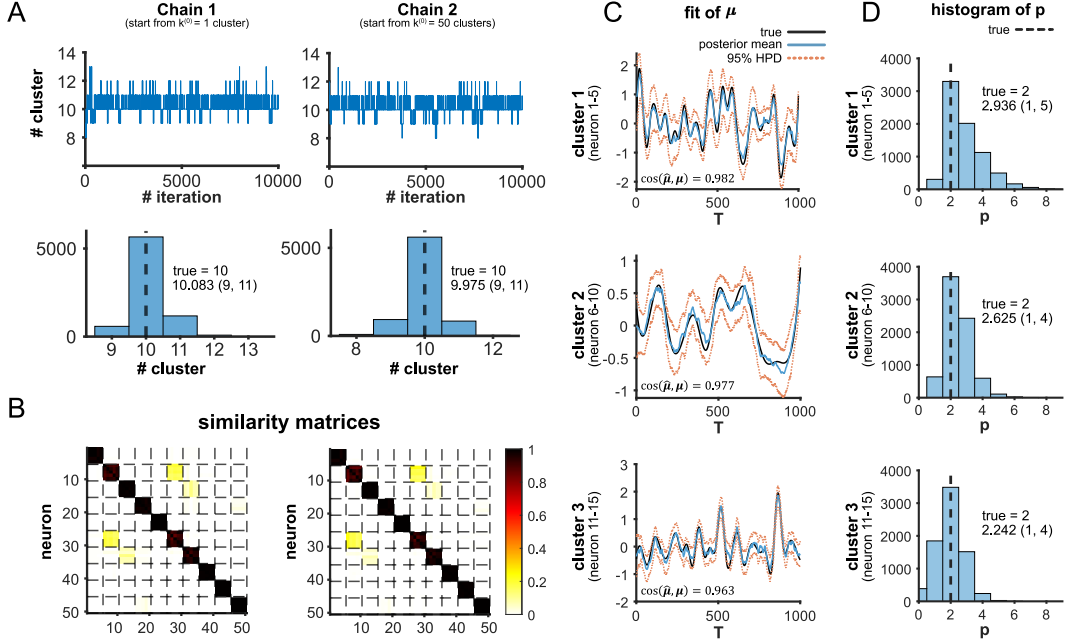}
	\caption{\textbf{Simulation Results.} 
 \textbf{A.} The trace plots and histograms of posterior samples for cluster number, initialized with all neurons as one cluster ($k^{(0)}=1$) and each single neuron as a separate cluster ($k^{(0)}=50$). The posterior means and 95\% HPD intervals of cluster number are also provided besides each histogram. \textbf{B.} The similarity matrices with neurons sorted by ground truth clustering structure. \textbf{C.} The inferred (colored) population baseline $\bm{\mu}^{(j)}$ to the first three detected clusters in Chain 1 (the cluster 1, 2 and 3 are what most neuron 1-5, 6-10, 11-15 belong to), where the black is the truth, the solid blue indicates the posterior mean and dashed orange lines represent the 95\% HPD regions. The cosine ("overlap") between true value $\bm{\mu}$ and posterior mean $\hat{\bm{\mu}}$ is shown at left-bottom corner for each cluster. \textbf{D.} The histograms of posterior samples of $p_j$, where dashed black lines are ground truths $p_j = 2$. The posterior means and 95\% HPD intervals are also provided.}
	\label{fig2}
\end{figure}

We then run an MCMC chain with 10,000 iterations for each simulation and initialize the chain with all neurons in just one cluster. In this single cluster, we initialize with $p_j=1$, $\bm{A}^{(j)} = 1$, $\bm{b}^{(j)} = 0$, $\bm{Q}^{(j)} = 0.01$. We initialize $\bm{\mu}^{(j)}$ and $\bm{X}^{(j)}$ in the same way as how we generate the data, by spline interpolation on 10 to 35 evenly spaced points, with corresponding value for each point drawn from $\mathcal{N}(0,0.5^2)$. The initial values for cluster-independent parameters are randomly sampled from $\delta_i\sim N(0,1)$ and $\bm{c}_i \sim \mathcal{N}(0,1)$. The sampling steps for model parameters ($\delta_i$,$\bm{c}_i$ and $\bm{\theta}^{(j)}$ without $p_j$) given $z_i$ and $p_j$ (Appendix \ref{xmu}-\ref{dynamics}) are repeated four times in each iteration. It takes about 456 minutes to run 10,000 iterations for one chain (the chain 1 shown in Figure \ref{fig2}), using a 3.40 GHz processor with 16 GB of RAM. 

Visual inspection of trace plots for different parameters shows that the MCMC chain converges generally after 2,500 iterations, and Geweke diagnostics \citep{Geweke1991} also show no convergence issues (all $p$-values $\geq 0.1$). For each MCMC chain, we summarize the MCMC results by presenting the posterior means and 95\% highest posterior density (HPD) regions for the unknown parameters, with the first 2,500 iterations discarded as the burn-in period.

To further evaluate whether our proposed MCMC algorithm can accurately recover the true values of the unknown model parameters, we compare the posterior summaries with the truth as follows. Let us denote $\zeta$ as the true value of any scalar parameter we are interested in exploring for our model, then the posterior mean of $\zeta$ in the $l$-th simulation is denoted as $\hat{\zeta}_{(l)}$, where $l=1, \cdots, 50$ and the 95\% HPD regions as $\mathcal{C}_{\zeta_{(l)}}$. The performance are evaluated by calculating 1) empirical MSE, $\text{MSE}_{\zeta} = \frac{1}{50}\sum_{l=1}^{50}\|\hat{\zeta}_{(l)} - \zeta\|_2^2$ and 2) empirical coverage probability (CP), $\text{CP}_{\zeta} = \frac{1}{50}\sum_{l=1}^{50} {\mathds 1}(\zeta \in \mathcal{C}_{\zeta_{(l)}})$, where $\mathds{1}(\cdot)$ represents the indicator function. For a $r$-dimensional vector parameter $\bm{\zeta} = (\zeta_1,\ldots,\zeta_r)'$, we calculate MSE and CP element-wise for each $\zeta_l$ for $l=1,\ldots,r$, and output the mean, minimum and maximum values among all $\text{MSE}_{\zeta_l}$ and $\text{CP}_{\zeta_l}$.

We find that both the number of clusters and the mixDPFA parameters $\bm{\theta}^{(j)}$ for each cluster can be accurately recovered. For instance, the true value of $k=10$, the $\text{MSE}_{k} = 0.0094$ and $\text{CP}_{k} = 1.00$ by averaging among 50 experiments. To save the space, we only display the inference performance for the two most important parameters, $\bm{\mu}^{(j)}$ and $p_j$, in Table~\ref{randSeedTab}. We summarize the results of $\bm{\mu}^{(j)}$ and $p_j$ according to clusters that the neuron most frequently belongs to. For example, Neuron 1 through Neuron 5 are most often assigned to the same cluster in our computation, and thus we name this cluster as ``neuron 1-5" to summarize the parameter results based on this cluster and similar interpretation is applied to the rest of cluster names in Table \ref{randSeedTab}. For $\bm{\mu}^{(j)}$, the typical element-wise mean value of MSE is around 0.01 to 0.05, which is pretty small; while the mean value of CP for the element-wise 95\% HPD interval are all above the nominal level 0.95. For latent dimension $p_j$, the average MSE among 50 random seeds ranges from 0.088 to 0.878, and all 95\% HPD intervals are all above the nominal level 0.95. All these results imply that we can accurately recover the true clustering assignments and corresponding model parameters for each cluster.

\begin{table*}
\caption{The MSE and coverage probability of inference on $\bm{\mu}=(\mu_1,\ldots,\mu_{T})'$ and $p_j$ for 50 random seeds, for clusters that most neuron 1-5, 6-10,..., 46-50 belong to. The column of $\text{MSE}_{\bm{\mu}^{(j)}} (\text{min, max})$ and $\text{CP}_{\bm{\mu}^{(j)}}(\text{min, max})$ shows the mean, minimum and maximum among all $\text{MSE}_{\mu_t^{(j)}}$ and $\text{CP}_{\mu_t^{(j)}}$, $t=1,\cdots,T=1000$}
\label{randSeedTab}
\begin{tabular}{@{}lrrrc@{}}
	\hline cluster 
	& \multicolumn{1}{c}{$\text{MSE}_{\bm{\mu}^{(j)}} (\text{min, max})$} & \multicolumn{1}{c} {$\text{CP}_{\bm{\mu}^{(j)}}(\text{min, max})$} & \multicolumn{1}{c}{$\text{MSE}_{p_j}$} & \multicolumn{1}{c}{$\text{CP}_{p_j}$}\\
	\hline
	neuron 1-5   & 0.031 (0.005, 0.313) &1.000 (0.920, 1.000) & 0.878 & 1.000\\
	neuron 6-10  & 0.009 (0.003, 0.129) & 1.000 (0.920, 1.000) & 0.538 & 1.000\\
	neuron 11-15 & 0.014 (0.005, 0.058) & 0.999 (0.960, 1.000) & 0.213 & 1.000\\
	neuron 16-20 & 0.009 (0.003, 0.049) & 1.000 (0.940, 1.000) & 0.504 & 1.000\\
	neuron 21-25 & 0.012 (0.003, 0.179) & 1.000 (0.960, 1.000) & 0.751 & 1.000\\
	neuron 26-30 & 0.020 (0.004, 1.082) & 0.997 (0.860, 1.000) & 0.179 & 1.000\\
	neuron 31-35 & 0.031 (0.006, 0.280) & 0.998 (0.820, 1.000) & 0.659 & 1.000\\
	neuron 36-40 & 0.052 (0.006, 0.271) & 0.975 (0.800, 1.000) & 0.224 & 1.000\\
	neuron 41-45 & 0.017 (0.004, 0.272) & 0.999 (0.880, 1.000) & 0.088 & 1.000\\
	neuron 46-50 & 0.042 (0.003, 0.241) & 0.985 (0.800, 1.000) & 0.287 & 1.000\\
	\hline
\end{tabular}
\end{table*}

To display the clustering performance in the figure much clearly, the results in Figure \ref{fig2} only use observations generated from one random seed. However, the performance for the other random seeds is similar. We run two independent chains with 10,000 iterations, where Chain 1 is initialized by assuming that all neurons are in one cluster and Chain 2 is initialized by assuming that every neuron is a unique cluster (50 clusters in total). The first 2,500 iterations are discarded as burn-in.

The trace plots for the number of clusters (Figure \ref{fig2}A) show that these two chains converge to the same ground truth ($10$ clusters), with prior $k \sim Geometric(\nu = 0.2)$, so $P(k\leq 10) \approx 0.9$. The analysis results are insensitive to the choice of $\nu$ (we have tried $\nu$ to 0.3 and 0.4, but they all produced similar results). We then evaluate similarity matrices (Figure \ref{fig2}B) where the entry $(i,l)$ is the posterior probability that data points $i$ and $l$ belong to the same cluster. The similarity matrices are sorted according to the the clustering structure generated from simulation, i.e., neuron 1-5, 6-10, 11-15,... belong to cluster 1, 2, 3, and so on. Therefore, if the algorithm can recover the simulated clustering structure, the diagonal blocks of similarity matrices will have high posterior probability (dark color in Figure \ref{fig2}B). The similarity matrices for these two chains show the same pattern. Namely, both chains recover the ground truth clustering assignments, but consistently show some slight confusions (light yellow shades) between cluster 2 (neuron 6-10) and cluster 6 (neuron 36-30), and cluster 3 (neuron 11-15) and cluster 7 (neuron 31-35). Overall, these results show that our method can accurately recover the ground truth number of clusters and the overall clustering structure as well as is insensitive to the initial number of clusters.

The mixDPFA parameters for each cluster are of particular interest for applications and interpretation in neuroscience. Thus, we further examine their inference performance in our sampling scheme.  In Figure \ref{fig2}, we show the inferred $\bm{\mu}^{(j)}$ and $p_j$ for the first three detected clusters (clusters that neurons 1-5, 6-10 and 11-15 most frequently belong to) in Chain 1. We evaluate the inference accuracy of $\bm{\mu}^{(j)}$ by checking the posterior means and 95\% highest posterior density (HPD) regions, which are showing accordingly using the blue solid line and red dash lines throughout the time $T$ in Figure \ref{fig2}C. We find that the posterior mean trajectory (the blue solid line) of $\bm{\mu}^{(j)}$ is very close to its true value (the black solid line), and additionally, the true value of $\bm{\mu}^{(j)}$ are fully covered by its 95\% HPD credible bands in the first two clusters with only a few points outside of its 95\% HPD credible bands for the third cluster. Moreover, we calculate the cosine ("overlap", used in \citet{Johnstone2009}) between posterior means $\bm{\hat{\mu}}^{(j)}$ and ground truths ${\bm{\mu}^{(j)}}$ by using ${\bm{\mu}^\top\bm{\hat{\mu}}}/{\|\bm{\mu}\|\|\bm{\hat{\mu}}\|}$ and the results are shown in Figure \ref{fig2}C, with the value closer to 1 the better. We then check the inference of latent dimension $p_j$ for each cluster. The histograms of posterior samples for $p_j$ (Figure \ref{fig2}D) show that we can recover the ground truth ($p_j = 2$) well. To further validate we can recover the different latent dimensions of $p_j$s, we generate an additional simulation using the same settings except for $p_j = 3$ for each cluster (see results in Appendix \ref{p3}).

\section{Multi-region Neural Spike Recordings}
\label{application}
We then apply the proposed clustering method to the Allen Institute Visual Coding Neuropixels dataset, a large, publicly available dataset for studying coding and signal propagation across cortical and thalamic visual areas. The dataset contains spiking activity from hundreds of neurons in multiple brain regions of an awake mouse. See detailed data description in \cite{Siegle2021}. Here we investigate the clustering structure of 83 neurons from four anatomical sites: 1) hippocampal CA1 (24 neurons), 2) dorsal part of the lateral geniculate complex (LGd, 36 neurons), 3) lateral posterior nucleus of the thalamus (LP, 12 neurons) and 4) primary visual cortex (VISp, 11 neurons). We analyze 50 second epochs ($T=500$ with bin size = 0.1 seconds) where neural activity was recording during different visual stimuli: two epochs where the visual stimulus was drifting gratings (D1 and D2), two epochs of spontaneous activity (S1 and S2), and one epoch with a natural movie stimulus (N). Only neurons with rates > 1Hz within the selected epochs are included and we analyze data with 100  millisecond bins. Since these neurons come from four brain regions, we might expect four clusters, and, to integrate this knowledge as prior information, we use a $k\sim Geometric(0.33)$, which makes $P(k\le4)=0.8$.

\begin{figure}[h!]
	\centering
	\includegraphics[width=1\textwidth]{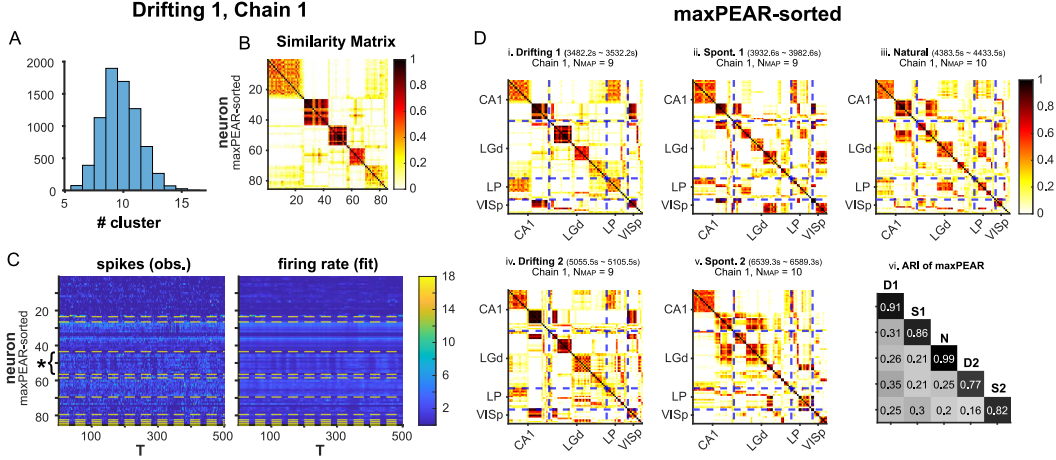}
	\caption{\textbf{Application in Neuropixels data.} \textbf{A.} The histogram of cluster numbers, in the first drifting grating epoch (Drifting 1, D1), using samples from iteration 2501 to 10,000. \textbf{B.} The posterior similarity matrix of Chain 1 in D1 epoch, sorted by the maxPEAR label. \textbf{C.} The observed spikes ($y_{it}$) and fitted mean firing rate ($\hat{\lambda}_{it}$, averaging from MCMC samples), sorted by the maxPEAR labels. The dashed yellow lines represent the maxPEAR-sorted neural populations. \textbf{D.} The posterior similarity matrices for 4 adjacent epochs and 1 further epoch with different visual stimuli, sorted according to maxPEAR estimate and anatomical label in D1 epoch. The last panel shows the ARI of the maxPEAR estimates. The diagonal is the ARI between two chains for the same data, while off-diagonal values show the average ARI (between two chains) of maxPEAR from two different epochs.}
	\label{fig3}
\end{figure}

For each data epoch, we run MCMC with 10,000 iterations. Because of low firing rates (and hence less information), the inferred clustering structure is more uncertain compared to the simulation results, and the average number of clusters is around 9 for epoch D1, for example, as shown in Figure \ref{fig3}A. To summarize the clustering results from posterior samples, we find a single estimate for cluster indices ${\hat{z}_i}$ by maximizing the posterior expected adjusted Rand index (maxPEAR, \citealt{Fritsch2009}). The maxPEAR estimates possess a shrinkage property, and it performs better compared to other estimating procedures, including procedures based on Maximum a posteriori (MAP), Blinder's loss \citep{Binder1978} and Dahl's criterion \citep{Dahl2006}. To show the overall clustering structures and the spiking patterns within each population, we present the maxPEAR-sorted posterior similarity matrix in Figure \ref{fig3}B (results sorted by MAP are similar, see Appendix \ref{supp_pixel}). We find that the latent trajectories accurately reconstruct the fluctuations in the observed spiking of each neuron for each of the inferred clusters as shown in Figure \ref{fig3}C, and the clusters have distinct patterns of activity that may be physiologically relevant. For instance, the maxPEAR-sorted cluster containing neuron 44-57 (marked with the asterisk in Figure \ref{fig3}C)  appears to contain neurons whose firing rate varies somewhat periodically over the course of the epoch, and these fluctuations are not present in the other maxPEAR-sorted clusters.

To examine the relationship between the clustering results and anatomy under D1, we additionally sort the neurons according to anatomical labels (Figure \ref{fig3}D-i). Although many identified clusters are neurons from the same anatomical area, clusters also include neurons from different regions and neurons within a region are often clustered into separate populations. For neurons clustered together, only 51\% of them belong to the same anatomical region, which implies that the mixture latent structure may more accurately represent the population of neurons compared to a simple assignment based on anatomy. 

Detailed interpretation of the inferred clusters and the trajectories of the latent variables is beyond the scope of this paper, but relating these clusters to features of the individual neurons and relating the latent trajectories to features of the visual stimulus have the potential to yield neuroscientific insights. For example, the clustering results for D1 suggest that there may be two clusters within area CA1: one which shares a latent state with neurons in area LP and a second which shares a latent state with neurons in area VISp. Comparing the composition of these clusters (e.g., excitatory vs. inhibitory neurons) and examining the latent trajectories underlying the clusters may show whether and how task variables are deferentially represented by these neurons.

To illustrate one possible type of comparison, we then evaluate the clustering patterns for five consecutive epochs with different visual stimuli (D1, S1, N, D2 and S2). We run two independent chains for each epoch, where the results from the second chain shown in Figure \ref{supp}A. The similarity matrices (Figure \ref{fig3}D) again show that the inferred clusters are partially but not directly related to the anatomical groups but with additional, substantial differences between the different visual stimuli. We also find that the clustering structure changes over time even for matched visual stimuli (D1 vs. D2 and S1 vs. S2). These changes may suggest that clusters are stimulus-dependent and can change over time. To quantify changes in clustering within and across epochs we evaluate the adjusted Rand index (ARI, \citet{Rand1971}) of maxPEAR estimates (Figure \ref{fig3}D-vi). We find that between-epoch comparisons tend to have lower similarity (average ARI from comparing 2 chains for each epoch) than within-epoch comparisons (different chains) for both maxPEAR and MAP (see Figure \ref{supp}C).

\section{Conclusion and Discussion}
\label{discussion}
In this paper, we have introduced a mixDPFA model to cluster neural spike trains using a Bayesian analysis, where the number of clusters and the dimension of latent states for the neural spike activities can be inferred directly from the data. Previous approaches to multi-population latent variable modeling have used anatomical information to label distinct groups of neurons, but this choice is somewhat arbitrary. Brain region and cell-type, for instance, can give contradictory population labels. The proposed mixDPFA method groups neurons by common latent factors, which may be useful for identifying "functional populations" of neurons. 

To design the computational algorithm for the proposed mixDPFA model, we have employed a partition-based algorithm from MFM to sample the number of clusters, while the dimension of latent states is inferred by a BDMCMC sampling. To efficiently sample cluster assignments ($z_i$) and latent dimension ($p_j$) in our model, we have approximated the marginalized likelihood (Equation \eqref{eq_marginal}) by a Poisson-Gamma conjugacy, which leads to a closed form for our sampling. This approximation might sacrifice somewhat accuracy of computing the marginalized likelihood. But since the sampling procedures of $p_j$ and $z_i$ depend on relative values of marginalized likelihood (c.f., Appendix \ref{Sample_A4} and Appendix \ref{Sample_A5}), this approximation does not substantially influence the sampling results here, either in simulations or in experimental data (similar discussions can be found in \cite{El-Sayyad1973} and \cite{Chan2009}).
Typically, this approximation works well for neural spiking data since the observation $y_{it}$ is small (the frequency is generally < 200 Hz), but it may not be accurate enough when $y_{it}$ is very large. More accurate methods may need to evaluate the marginalized likelihood when applying this method to count data with large value observations.

Although the proposed method can describe data and cluster neural spiking activity successfully, there are some potential improvements. First, to make the model identifiable, here we put diagonal constraints on $\bm{A}^{(j)}$ and $\bm{Q}^{(j)}$ and constrain $\bm{\mu}^{(j)}$ and $\bm{X}^{(j)}$ to have mean zero. The assumption that $\bm{A}^{(j)}$ and $\bm{Q}^{(j)}$ are diagonal does not allow interaction between latent factors. However, these interactions could be allowed by instead constraining $\bm{X}'^{(j)}\bm{X}^{(j)}$ to be diagonal \citep{Krzanowski1994a,Krzanowski1994b,Fokoue2003}. Such a constraint could allow unique solutions for the (P)LDS and GPFA. Second, a deterministic approximation of MCMC, such as variational inference (VI) may be more computationally efficient. Standard methods for fitting the PLDS could be used directly in the VI updates, and if we further use a stick-breaking representation for the MFM model, it would be straightforward to use VI for clustering as well, similar to \cite{Blei2006}. Third, the neurons may change states along the time, when the stimulus or internal states changes. To consider the state changes when clustering neurons, we can further extend the current model to be a mixture of the Poisson switching-LDS \citep{Fox2009,Murphy2012}, which leads to a double layer mixture model.  

To sum up, as the number of neurons and brain regions that neuroscientists are able to record simultaneously continues to grow, understanding the latent structure of multiple populations will be a major statistical challenge. The Bayesian approach to clustering neural spike activities introduced here converges fast and is insensitive to the initial cluster assignments, and may, thus, be a useful tool for identifying "functional populations" of neurons.

\begin{appendix}
\section{MCMC updates}\label{MCMC}
The joint posterior distribution for parameters in mixDPFA is shown in Equation \eqref{FullPost}. The posteriors are sampled by a MCMC sampler. In each iteration, the sampling scheme has three key steps: (1) sample the model parameters assuming the labels $z_i$s and the number of latent factors $p_j$s are given, 
(2) sample the number of latent factors $p_j$s given the rest known and 
(3) sample the cluster indices $z_i$s given model parameters. In our sampling, the key step (1) have been repeated several times (four times in the our simulation and application), and is first conducted without considering constraints for $\mu_t^{(j)}$ and $\bm x_t^{(j)}$, and then project the samples onto the constraint space for $\sum_{t=1}^{T}\mu_t^{(j)} = 0$ and $\sum_{t=1}^{T}\bm{x}^{(j)}_t = \bm{0}$, the same strategy has been used in \citet{https://doi.org/10.48550/arxiv.1812.05741}. The key step (1) further can be decomposed into three sub-steps, i.e., updating 1) population baseline $\bm{\mu}^{(j)}$ and latent factors $\bm{X}^{(j)}$; 2) cluster-independent $\delta_i$ and $\bm{c}_i$; and 3) parameters governing linear dynamics $\{h^{(j)}, g^{(j)}, (\sigma^2)^{(j)}, \bm{A}^{(j)}, \bm{b}^{(j)}, \bm{Q}^{(j)}\}$. We present the detailed sampling schemes from Step \ref{Sample_A1} to Step \ref{Sample_A5} below.

\subsection{Update population baseline and latent factors}\label{Sample_A1}
\label{xmu}
In this section, we provide details of sampling algorithms to draw the latent states $\bm{X}^{(j)}$ and population baseline $\bm{\mu}^{(j)}$ from the full conditional distribution. In order to explain the algorithm in a clearer way, we just focus on sampling in one cluster, and thus in this section we suspend the superscript $(j)$ or $j$ as the cluster index in our notations.

Assume there are $n$ neurons in the given cluster. Then, $y_{it} \sim Poi(\lambda_{it})$, $\log \lambda_{it} = \delta_i + \widetilde{\bm{c}}'_i\widetilde{\bm{x}}_t$ 
for $i=1,\ldots, n$ and $t=1,\ldots,T$, where $\widetilde{\bm{c}}_i = (1, \bm{c}_i')'$ and 
$\widetilde{\bm{x}}_t = \left(\mu_t, \bm{x}_t'\right)'$. According to Equation \eqref{stateEquation}, the $\widetilde{\bm{x}}_t$ follows linear dynamics $\widetilde{\bm{x}}_{t+1}| \widetilde{\bm{x}}_{t} \sim \mathcal{N}(\widetilde{\bm{A}}\widetilde{\bm{x}}_{t} + \widetilde{\bm{b}}, \widetilde{\bm{Q}})$, where $\widetilde{\bm{A}}$, $\widetilde{\bm{b}}$ and $\widetilde{\bm{Q}}$ are defined in section \ref{infer}. We assume the prior for $\widetilde{\bm{x}}_{1}$ is $\widetilde{\bm{x}}_{1} \sim \mathcal{N}(\bm{0}, \bm{I}_{p+1})$. 
If we further denote $\widetilde{\bm{y}}_t = \left(y_{1t},\ldots,y_{nt}\right)'$, $\widetilde{\bm{\lambda}}_t = \left(y_{1t},\ldots,y_{nt}\right)'$, $\bm{\delta} = (\delta_1,\ldots,\delta_n)'$ and $\widetilde{\bm{C}} = \left(\widetilde{\bm{c}}_1,\ldots,\widetilde{\bm{c}}_n\right)'$, the full conditional distribution of $\widetilde{\bm{X}}=(\bm{\mu}, \bm{X})$ is 
\begin{eqnarray}
    && P(\widetilde{\bm{X}}\mid \{\widetilde{\bm{y}}_t\}_{t=1}^T, \bm{\delta}, \widetilde{\bm{C}}, \widetilde{\bm{A}}, \widetilde{\bm{b}}, \widetilde{\bm{Q}})\nonumber\\
    &\propto& \exp{\left(\sum_{t=1}^{T}\widetilde{\bm{y}}'_t(\bm{\delta} + \widetilde{\bm{C}}\widetilde{\bm{x}}_t) - \bm{1}'_n\widetilde{\bm{\lambda}}_t\right)}\exp{\left(-\frac{1}{2}\| \widetilde{\bm{x}}_1\|_2^2\right)}\prod_{t=2}^{T}\exp{\left(-\frac{1}{2}\bm{s}'_t|\widetilde{\bm{Q}}|^{-1}\bm{s}_t\right)}, \label{eqn_cd_xu}
\end{eqnarray}
where $\bm{s}_t = \widetilde{\bm{x}}_t - \widetilde{\bm{A}}\widetilde{\bm{x}}_{t-1} - \widetilde{\bm{b}}$. However, this full conditional distribution 
above 
has no closed form, thus we could not use Gibbs sampling in this step. Instead, we sampling $\widetilde{\bm{X}}$ with two key parts as following and the detailed algorithm is summarized in Algorithm \ref{algo}.

\subsubsection{P\'olya-Gamma Augmentation}
Although the mixDPFA model does not directly follow the PG augmentation scheme \citep{doi:10.1080/01621459.2013.829001}, we can approximate the Poisson distribution by a negative binomial (NB) distribution, i.e., $\lim_{r\to \infty} \text{NB}(r, \sigma(\psi - \log r)) = \text{Poisson}(e^{\psi})$, where $\sigma(\psi) = e^{\psi}/(1 + e^{\psi})$ and $\text{NB}(r,p)$ denotes the NB distribution with $rp/(1-p)$ as its expectation. We approximate the mixDPFA model using a NB, then we instead sample a mixture of NB factor analyzers using the PG scheme \citep{Windle2013}. Further, 
we use the forward-filtering-backward-sampling (FFBS) algorithm \citep{10.1093/biomet/81.3.541, FrhwirthSchnatter1994DataAA} to update $\widetilde{\bm{X}}$ (i.e., the latent states $\bm{X}$ and population baseline $\bm{\mu}$). These two steps are summarized in Step 1 and Step 2 of Algorithm \ref{algo}. 

\subsubsection{Metropolis-Hastings Step}
The samples of $\widetilde{\bm{X}}$ yielded from FFBS algorithm is just a proposal, where we employ a Metropolis-Hastings (MH) step to reject or accept this proposal. 
In this step, the dispersion parameter $r$ in NB distribution becomes a tuning parameter, to balance acceptance rate and autocorrelation in MH. When $r$ is large, the approximation to Poisson observation is accurate and the MH performs similar to the Gibbs sampler. The neurons at different time points can have unique tuning parameters $r_{it}$, although in the implementation of this paper we set them as a constant $r_{it} = r$. The MH step is summarized in Step 3 of Algorithm \ref{algo}.

\RestyleAlgo{ruled}
\begin{algorithm}[!hpt]
	\caption{P\'olya-Gamma-Metropolis-Hastings Algorithm (PG-MH) for Poisson Dynamic Model}
	\label{algo}
	Given the sample from the $(G-1)$-th iteration $\widetilde{\bm{x}}^{(G-1)}_{ t}$ and $\bm{U} = \{\{\widetilde{\bm{c}}_i, \delta_i\}_{i=1}^{n}, \widetilde{\bm{A}}, \widetilde{\bm{b}}, \widetilde{\bm{Q}}\}$.\\
	\;
	\textbf{1.} Sample $\omega_{it}$ from PG distribution and calculate $\hat{y}_{it}$, which follows $\mathcal{N}(\widetilde{\bm{c}}'_i\widetilde{\bm{x}}_t^{(G-1)}, \omega_{it}^{-1})$;\\
	\For{$t = 1,\ldots, T$}{
		\For{$i = 1,\ldots, n$}{
			sample $\omega_{it} \sim PG(r_{it} + y_{it}, \delta_i + \widetilde{\bm{c}}'_i\widetilde{\bm{x}}_t^{(G-1)} - \log r_{it})$\\
			$\kappa_{it} = (y_{it} - r_{it})/2 + \omega_{it}(\log r_{it} - \delta_i)$\\
			$\hat{y}_{it} = \omega_{it}^{-1}\kappa_{it}$
	}}
	\;
	\textbf{2.} Forward-filtering-backward-sampling (FFBS) for $\widetilde{\bm{X}}$;\\
	denote $\hat{\bm{y}}_t = (\hat{y}_{1t},\ldots,\hat{y}_{Nt})'$, $\bm{\Omega}_t = Diag([\omega_{1t},\ldots,\omega_{Nt}])$ and $\widetilde{\bm{C}} = (\widetilde{\bm{c}}_1,\ldots,\widetilde{\bm{c}}_N)'$;\\
	\For{$t=1,\ldots, T$}{
		$\bm{m}_{t|t-1} = \widetilde{\bm{A}}\bm{m}_{t-1} + \widetilde{\bm{b}}$\\
		$\bm{V}_{t|t-1} = \widetilde{\bm{A}}\bm{V}_{t-1}\widetilde{\bm{A}}' + \widetilde{\bm{Q}}$\\
		$\bm{K}_t = \bm{V}_{t|t-1}\widetilde{\bm{C}}'(\widetilde{\bm{C}}\bm{V}_{t|t-1}\widetilde{\bm{C}}' + \bm{\Omega}_t^{-1})^{-1}$\\
		$\bm{m}_t = \bm{m}_{t|t-1} + \bm{K}_t(\hat{\bm{y}}_t - \widetilde{\bm{C}}\bm{m}_{t|t-1})$\\
		$\bm{V}_t = (\bm{I} - \bm{K}_t\widetilde{\bm{C}})\bm{V}_{t|t-1}$
	}
	sample $\widetilde{\bm{x}}^*_T \sim \mathcal{N}(\bm{m}_T, \bm{V}_T)$;\\
	\For{$t = T-1,\ldots,1$}{
		$\bm{J}_t = \bm{V}_t\widetilde{\bm{A}}'(\widetilde{\bm{A}}\bm{V}_t\widetilde{\bm{A}}' + \widetilde{\bm{Q}})^{-1}$\\
		$\bm{m}^*_t = \bm{m}_t + \bm{J}_t(\bm{x}^*_{t+1} - \widetilde{\bm{A}}\bm{m}_t - \widetilde{\bm{b}})$\\
		$\bm{V}^*_t = (\bm{I} - \bm{J}_t\widetilde{\bm{A}})\bm{V}_t$\\
		sample 	$\bm{x}^*_t \sim \mathcal{N}(\bm{m}^*_t, \bm{V}^*_t)$
	}
	\;
	\textbf{3.} Accept or reject the proposal $\widetilde{\bm{X}}^*$ by computing the acceptance ratio:\\
	\begin{align*}
		\zeta &= \frac{\pi(\widetilde{\bm{X}}^*|\{\bm{y}_i\}_{i=1}^n, \bm{U})}{\pi(\widetilde{\bm{X}}^{(G-1)}|\{\bm{y}_i\}_{i=1}^n, \bm{U})}\frac{q(\widetilde{\bm{X}}^{(G-1)}|\widetilde{\bm{X}}^*, \bm{U})}{q(\widetilde{\bm{X}}^*|\widetilde{\bm{X}}^{(G-1)}, \bm{U})}\\
		&= \frac{P(\{\bm{y}_i\}_{i=1}^n|\widetilde{\bm{X}}^*)}{P(\{\bm{y}_i\}_{i=1}^n|\widetilde{\bm{X}}^{(G-1)})}\frac{NB(\{\bm{y}_i\}_{i=1}^n|\widetilde{\bm{X}}^{(G-1)}, \bm{R})}{NB(\{\bm{y}_i\}_{i=1}^n|\widetilde{\bm{X}}^*, \bm{R})},
	\end{align*}
	where $\bm{R} = \{r_{it}\}$ is matrix for dispersion parameters for each neuron at all time points. $P(\cdot)$ denotes the Poisson likelihood, and $NB(\cdot)$ denotes the negative binomial likelihood; and 
	accept the proposal $\widetilde{\bm{X}}^*$ with probability $\text{min}(1, \zeta)$.
\end{algorithm}

\subsection{Update neuron-specific baseline and loading}\label{Sample_A2}
\label{deltac}
From the matrix representation of mixDPFA in \eqref{eq_matrix}, i.e., 
$\log\bm{\lambda}_i = \delta_i\bm{1}_T + \bm{\mu}^{(j)}+ \bm{X}^{(j)}\bm{c}_i$, it is easy to see that given $\bm{\mu}^{(j)}$ and $\bm{X}^{(j)}$ are known, the update of $\delta_i$ and $\bm{c}_i$ is just a regular Bayesian Poisson regression problem. Thus, we can sample the full conditional distribution of $\delta_i$ and $\bm c_i$ by a Hamiltonian Monte-Carlo (HMC, \citet{Duane1987,Neal1994}). This is implemented by calling \verb+hmcSampler+ and \verb+drawSamples+ functions in MATLAB.

\subsection{Update parameters of latent state}\label{Sample_A3}
\label{dynamics}
Here we provide the update for parameters governing linear dynamics, i.e., $h^{(j)}$, $g^{(j)}$, $(\sigma^{2})^{(j)}$, $\bm{A}^{(j)}$, $\bm{b}^{(j)}$ and $\bm{Q}^{(j)}$.  Since we assume $\bm{A}^{(j)} = diag(a_1^{(j)},\ldots, a_p^{(j)})$ and $\bm{Q}^{(j)} = diag(q_1^{(j)},\ldots, q_p^{(j)})$, we can update $\bm{A}^{(j)}$, $\bm{b}^{(j)}$ and $\bm{Q}^{(j)}$ for each diagonal element separately, as these updates in $h^{(j)}$, $g^{(j)}$ and $(\sigma^{2})^{(j)}$. Here, we update $h^{(j)}$, $g^{(j)}$ and $\sigma^{2(j)}$ as follows.

Let us denote $\bm{\mu}_{2:T}^{(j)} = \left(\mu_2^{(j)}, \ldots, \mu_{T}^{(j)}\right)'$ and $\widetilde{\bm{\mu}}_{1:(T-1)}^{(j)} = \left(\bm{1}_{T-1}, \bm{\mu}_{1:(T-1)}^{(j)}\right)$, with $\bm{\mu}_{1:(T-1)}^{(j)} = \left(\mu_1^{(j)}, \ldots, \mu_{T-1}^{(j)}\right)'$. The full conditional distributions for $\sigma^{2(j)}$ and $\left( g^{(j)}, h^{(j)}\right)'$ are:
\begin{align*}
	(\sigma^{2})^{(j)}|\{\mu_t^{(j)}\}_{t=1}^T &\sim IG\left(\frac{\nu_0 + T-1}{2}, \frac{\nu_0\sigma^2_0 + \bm{\mu}_{2:T}'^{(j)}\bm{\mu}_{2:T}^{(j)} + \bm{\tau}'_0\bm{\Lambda}_0\bm{\tau}_0 - \bm{\tau}'_n\bm{\Lambda}_n\bm{\tau}_n }{2}\right),\\
	\left( g^{(j)}, h^{(j)}\right)'|\{\mu_t^{(j)}\}_{t=1}^T &\sim \mathcal{N}(\bm{\tau}_n, (\sigma^{2})^{(j)}\bm{\Lambda}_n^{-1}),
\end{align*}
with $\bm{\Lambda}_n = \widetilde{\bm{\mu}}_{1:(T-1)}'^{(j)}\widetilde{\bm{\mu}}_{1:(T-1)}^{(j)} + \bm{\Lambda}_0$, and $\bm{\tau}_n = \bm{\Lambda}_n^{-1}\left(\widetilde{\bm{\mu}}_{1:(T-1)}'^{(j)}\bm{\mu}_{2:T}^{(j)} + \bm{\Lambda}_0\bm{\tau}_0\right)$.

\subsection{Update number of latent factors}\label{Sample_A4}
\label{p}
The number of latent factors for each cluster $p_j$ is sampled by a birth-death MCMC method (BDMCMC, \cite{10.1214/aos/1016120364}) as elaborated in Algorithm \ref{bdmcmc}, similar to \cite{Fokoue2003}. 
In Algorithm \ref{bdmcmc}, we set the birth rate as $\beta = 0.5$ and duration time as $\rho = 1$. Define $\mathcal{C} = \{\bm{X}^{(j)}_{*1}, \bm{X}^{(j)}_{*2},...\}$, where $\bm{X}^{(j)}_{*l}$ denotes the $l$-th column of $\bm{X}^{(j)}$. Let $\mathcal{C}\cup \{\bm{X}^{(j)}_{*l}\}$ and $\mathcal{C}\backslash \{\bm{X}^{(j)}_{*l}\}$ denote the addition and deletion of column $\bm{X}^{(j)}_{*l}$ from the current configuration $\mathcal{C}$. Further, denote $M(\cdot)$ as the marginalized likelihood by integrating $\bm{c}_i \sim \mathcal{N}(\bm{0}, \bm{I}_{p_j})$ out. The resulting sampling scheme is summarized in Algorithm \ref{bdmcmc}.

\RestyleAlgo{ruled}
\begin{algorithm}[!hpt]
	\caption{Birth-death point process for $p_j$}
	\label{bdmcmc}
	In the iteration $G$, set the "elapsed birth-death time" $t_{fa} = 0$ and $p_j = p_j^{(G-1)}$\\
	\Repeat{$t_{fa} \geq \rho$}{
		Compute $\xi_k(\mathcal{C}) = \frac{M(\mathcal{C}\backslash \{\bm{X}^{(j)}_{*k}\})}{M(\mathcal{C})}\frac{\beta}{\alpha}$\\
		Compute $\xi(\mathcal{C}) = \sum_{k=1}^{p_j} = \xi_k(\mathcal{C})$\\
		Simulate $s \sim \text{Exp}(1/(\beta + \xi(\mathcal{C})))$ and set $t_{fa} = t_{fa} + s$\\
		\uIf{$\text{Bern}(\beta/(\beta + \xi(\mathcal{C}))) = 1$}{
			Set $q = q+1$\\
			Simulate $\bm{X}^{(j)}_{*q}$ from prior $\bm{H}$\\
			Set $\mathcal{C} = \mathcal{C} \cup \{\bm{X}^{(j)}_{*q}\}$
		}
		\Else{
			Simulate $k' = \text{Mn}(\xi_1(\mathcal{C})/\xi(\mathcal{C}),...,\xi_q(\mathcal{C})/\xi(\mathcal{C}) )$\\
			Set $\mathcal{C} = \mathcal{C}\backslash \{\bm{X}^{(j)}_{*k'}\}$\\
			Set $q = q - 1$
		}
	}
\end{algorithm}

\subsection{Update labels}\label{Sample_A5}
\label{zi}
To update the cluster labels $z_i$s, we use a partition based algorithm, similarly as described in \cite{Miller2018}. Let $\mathcal{C}$ denote a partition of neurons, and $\mathcal{C} \backslash i$ denote the partition obtained by removing neuron $i$ from $\mathcal{C}$.
\begin{enumerate}
	\item Initialize $\mathcal{C}$ and $\{\bm{\theta}^{(c)}: c\in\mathcal{C}\}$ (e.g., one cluster for all neurons in our simulation).
	\item Repeat the following Step a) and Step b) $G$ times to obtain $G$ samples. For $i=1,\ldots,N$: remove neuron $i$ from $\mathcal{C}$ and place it:
	\begin{enumerate}
		\item in $c\in\mathcal{C}\backslash i$ with probability $\propto (|c| +\gamma)M_{\bm{\theta}^{(c)}}(\bm{y}_i)$, where $\gamma$ is the hyperparameter of the Dirichlet distribution in Equation \eqref{MFMM} and $M_{\bm{\theta}^{(c)}}(\bm{y}_i)$ denotes the marginalized likelihood of neuron $i$ in cluster $c$, when integrating the loading $\bm{c}_i$ out.
		\item in a new cluster $c^*$ with probability $\propto \gamma \frac{V_n(s+1)}{V_n(s)}M_{\bm{\theta}^{(c^*)}}(\bm{y}_i)$, where $s$ is the number of partitions by removing the neuron $i$ and $V_n(s) = \sum_{l=1}^{\infty}\frac{l_{(s)}}{(\gamma l)^{(n)}}f_k(l)$, with $x^{(m)} = x(x+1)\cdots(x+m-1)$, $x_{(m)} = x(x-1)\cdots(x-m+1)$, $x^{(0)} = 1$ and $x_{(0)} = 1$.
	\end{enumerate}
\end{enumerate}
When evaluating the likelihood, we marginalize the cluster-independent loading $\bm{c}_i$ out. This is necessary for the high dimensional situation, otherwise the chain will stop moving.

One issue with incremental Gibbs samplers such as Algorithm 3 and 8 in \citet{Neal2000}, when applied to DPM, is that mixing can be somewhat slow. To further improve the mixing, we intersperse the "split-merge" MH updates \citep{10.1214/07-BA219,doi:10.1198/1061860043001} between Gibbs sweeps, 
as in \citep{Miller2018}.
After specifying the initial values for unknown parameters in the mixDPFA models, we then can start to run our MCMC sampling from {\it Step \ref{Sample_A1}} to {\it Step \ref{Sample_A5}} until the MCMC has converged. The convergence has been evaluated informally by looking at trace plots. 

\section{Simulation with $p_j=3$ for each cluster}
\label{p3}
Here we generate another set neural spikes for 50 neurons (10 clusters and 5 neurons in each cluster), using the same settings as in Section \ref{sim} except that now $p_j=3$ for each cluster. We have run a MCMC for 10,000 iterations discarding the first 2,500 iterations as a burn-in period and the results are shown below in Figure \ref{supp_p3}. 

\begin{figure}[h!]
	\centering
	\includegraphics[width=1\textwidth]{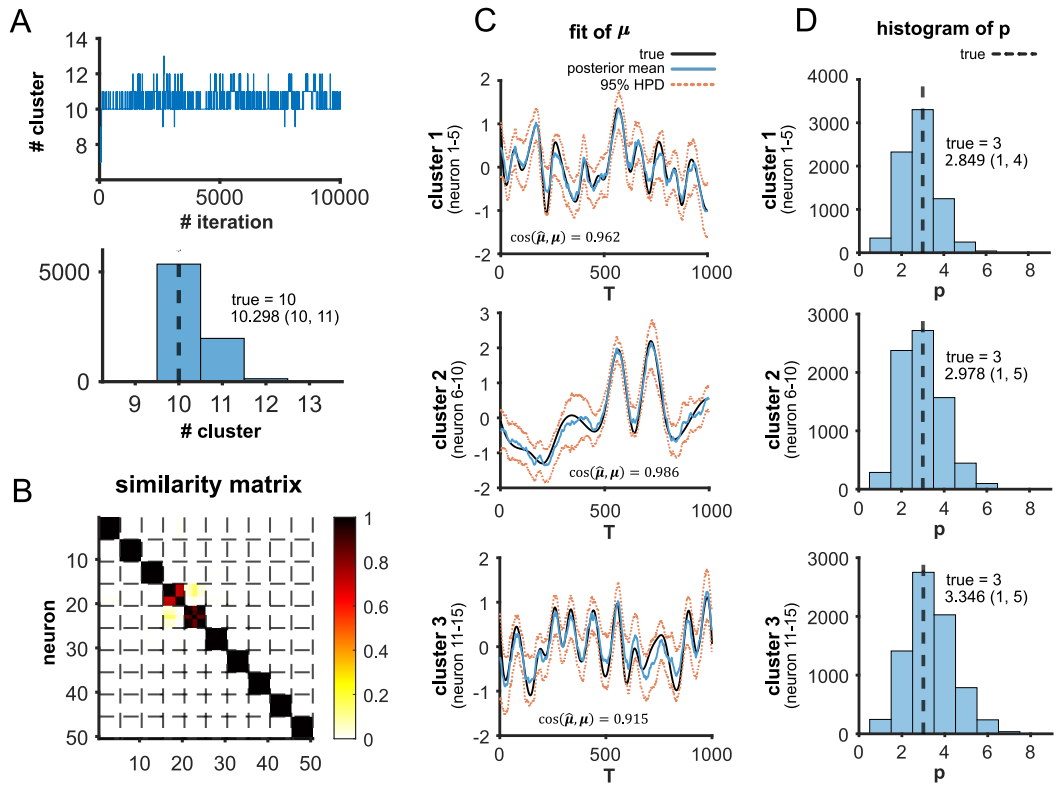}
	\caption{\textbf{Simulation Results with $p_j=3$ for each cluster.} \textbf{A.} The trace plot and histogram for posterior samples of cluster number with $k^{(0)}=1$. The posterior mean and 95\% HPD interval of the cluster numbers are provided. \textbf{B.} The similarity matrices with neurons sorted by ground truth clustering structure. \textbf{C.} The inferred mixDPFA parameters corresponding to the first three detected clusters are displayed. The true (black) and the inferred (colored) population baseline $\bm{\mu}^{(j)}$. In colored lines, the solid blue are posterior means and dashed orange represent the 95\% HPD regions. \textbf{D.} The histograms of posterior samples of $p_j$, where dashed black lines are ground truths $p_j = 3$. The posterior means and 95\% HPD intervals are also provided.}
	\label{supp_p3}
\end{figure}

\section{Supplementary Results for Neuropixels Application}
\label{supp_pixel}
\begin{figure}[h!]
	\centering
	\includegraphics[width=1\textwidth]{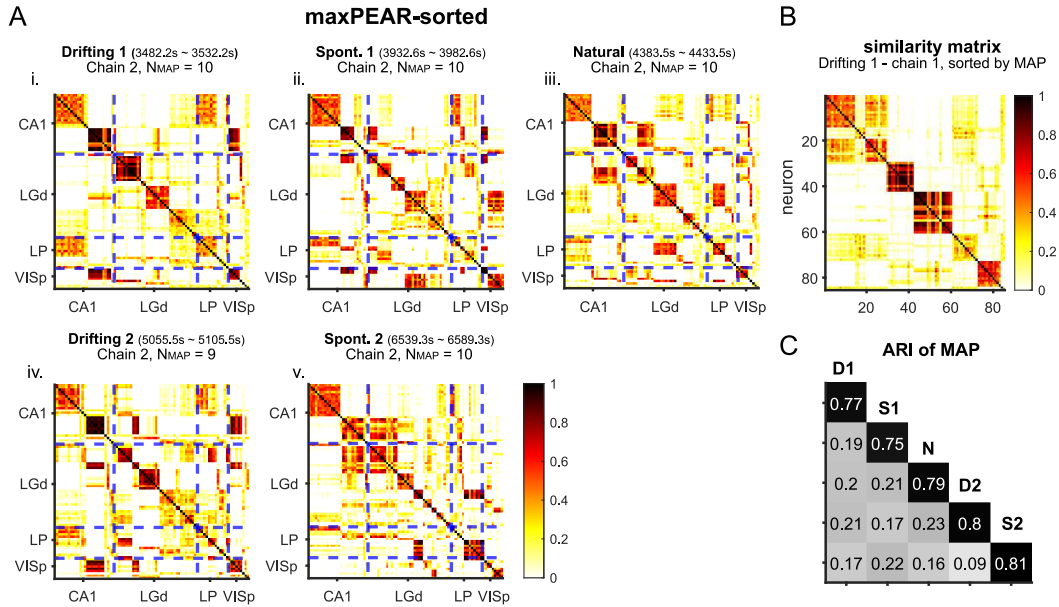}
	\caption{\textbf{Supplementary Results for Neuropixels Application.} \textbf{A.} Posterior similarity matrices for the second chains. Neurons are sorted as in the first panel of Figure \ref{fig3}D. \textbf{B.} The posterior similarity matrix sorted by MAP, for chain 1 when the mouse is exposed to drifting gratings as in Figure \ref{fig3}. \textbf{C.}  ARI of MAP estimates. The diagonal is ARI between 2 chains, and the off-diagonal is mean ARI of MAP for 4 combinations between two chains from two different epochs.}
	\label{supp}
\end{figure}

This section, we show supplementary results when applying the proposed model to the Neuropixels dataset (Multi-region neural spike recordings). Here, we show 1) clustering results in another independent chain for each epoch (Figure \ref{supp}A), with neurons sorted as in the first panel of Figure \ref{fig3}D and 2) results sorted by maximum a posteriori probability (MAP) estimates (Figure \ref{supp}B and C)

\end{appendix}

\begin{acks}[Acknowledgments]
The authors would like to thank the anonymous referees, the Associate
Editors and the Editors for the constructive comments to improve the
quality of this paper. Ms.Wei wants to acknowledge the National Science Foundation (NSF) Grant No. 1931249 for supporting his Ph.D. study. Dr. Stevenson's research has also been supported by the NSF Grant No. 1931249. 
In addition, Dr. Wang is grateful for the NSF Grant No. 1848451 to support her research. 
\end{acks}



\bibliographystyle{imsart-nameyear} 
\bibliography{MyCollection}       

\end{document}